\documentclass[11pt,twoside]{atmp}

\usepackage{amsmath,amssymb}

\usepackage{color}

\begin{document}

\title[Axial Current and  ....]
{Axial Current and   Noether Charge}


\author[Prasanta Mahato]{Prasanta Mahato}

\address{Narasinha Dutt College,
         Howrah, West Bengal, India 711 101}  
\addressemail{pmahato@dataone.in}

\begin{abstract}

A decade ago, a Lagrangian density  has been proposed by the author where only the local symmetries of the Lorentz subgroup of (A)ds group is retained. This formalism has been found to produce some results encompassing that of standard Einstein-Hilbert formalism.   In  the present article, the conserved axial vector matter currents, constructed in some earlier paper, have been found to be a result of  Noether's theorem.
  
 
\end{abstract}
\maketitle
\section{Introduction} Kibble\cite{Kib61} and Sciama\cite{Sci62} have pointed out that the \textit{Poincar$\acute{e}$}  group, which is the semi-direct product of translation and Lorentz rotation, is the underlying gauge group of gravity and found the so-called Einstein-Cartan theory  where mass-energy of matter is related to the curvature and spin of matter is related to the torsion of space-time.
 From geometrical point of view there is an important connection between the de Sitter group and the \textit{Poincar$\acute{e}$} group. It is a well known fact that the \textit{Poincar$\acute{e}$} group can be obtained from the  de Sitter group by an appropriate \textit{In$\ddot{o}$n$\ddot{u}$-Wigner} contraction\cite{Ino53,Dre85}. 
In the late 1970s, MacDowell and Mansouri\cite{Mac77} have introduced a
new approach of gravity, based on broken symmetry in a type of gauge theory. Macdowell-Mansouri gravity is based on the (A)dS group which retains only the local symmetries of the Lorentz subgroup. Even at the level of the action, the exact local (A)dS symmetry is clearly broken, whereas local Lorentz symmetry is retained.

 A decade ago,  a gravitaty Lagrangian has been
proposed\cite{Mah02a} where a
 Lorentz invariant part of the de Sitter Pontryagin density has been treated as
 the Einstein-Hilbert Lagrangian.  In this approach the role of torsion in the underlying manifold
is multiplicative
   rather than additive one and  the  Lagrangian  looks like
    $\mathbf{torsion \otimes curvature}$.   This indicates that torsion is uniformly nonzero
   everywhere. In the geometrical sense, this implies that
   microlocal space-time is such that at every point there is a
   direction vector (vortex line) attached to it. This effectively
   corresponds to the non commutative geometry having the manifold
   $M_{4}\times Z_{2}$, where the discrete space $Z_{2}$ is just
   not the two point space\cite{Con94}, but appears as an attached direction vector. Considering torsion and torsion-less
  connection as independent fields\cite{Mah04}, it has been found that $\kappa$  of Einstein-Hilbert Lagrangian, appears as an integration constant in such a way that it is linked with the topological Nieh-Yan density of $U_{4}$ space.
  If we consider axial vector torsion together with a scalar field $\phi$ connected to a local  scale factor\cite{Mah05}, then the Euler-Lagrange equations not only give the constancy of the gravitational constant but they also link, in laboratory scale, the mass of the scalar field with the Nieh-Yan density and, in cosmic scale of Friedmann-Robertson-Walker(FRW) cosmology, they predict only three kinds of the phenomenological energy density representing mass, radiation and cosmological constant. In another paper\cite{Mah07a}, it has been shown that this scalar field may also be interpreted to be linked with the dark matter and dark radiation. Also it has been shown that, using field equations of all fields except the frame field, the starting Lagrangian  reduces to a generic $f(\mathcal{R})$ gravity Lagrangian  which, for FRW metric,  gives standard FRW cosmology\cite{Mah07b}. Here $\mathcal{R}$ is the Ricci scalar of the Riemann curvature tensor. But for non-FRW metric, in particular of reference\cite{Lob07a}, with some particular choice of the functions of the scalar field $\phi$ one gets $f(\mathcal{R})=f_0\mathcal{R}^{1+v^2_{tg}}$, where $v_{tg}$ is the constant tangential velocity of the stars and gas clouds in circular orbits in the outskirts of spiral galaxies. With this choice of functions of $\phi$ no dark matter is required to explain flat galactic  rotation curves. This does not contradict with other  $f(\mathcal{R})$ type
gravity theories, which in general generate bifurcations\cite{Bifurcation} and anisotropy
of the universe\cite{Vig11}. In a recent paper\cite{Mah09}, we see that variation of torsion in the action gives us the axial vector 1-form $j_5=\overline{\Psi}\gamma_5\gamma\Psi$ to be an exact form. If we consider FRW geometry to be in the background then the FRW postulate\cite{Mah09},  makes it possible to define an axial vector current 3-form $	J^A_1$ as a product of torsion and matter current $j_5$. $	J^A_1$ is conserved and as well as gauge invariant. In  manifolds having arbitrary background geometry the product of $j_5$ and an exact 2-form $F$ gives us another gauge invariant conserved current $	J^A_2$. These conserved axial currents implies pseudoscalar conserved charges. \textbf{\textit{In this article we are going to study the possible connection of the conserved current $	J^A_2$ with a Noether symmetry in the presence of an electro-magnetic field or a Killing vector field. In the following  section  we   intend to show that the conserved axial vector matter current $	J^A_2$ of reference\cite{Mah09}  is a consequence of  Noether's theorem.}} The last section contains discussion.



\noindent

		
\section{Spinorial Matter and Axial Noether  Current}
We may define an axial vector current 3-form\cite{Mah09}, given by
\begin{eqnarray}
	J^A_{EM} \equiv	
\kappa(\overline{\Psi}\gamma_5\gamma\Psi\wedge F)\label{eqn:abc147}
	\end{eqnarray}
	where $F=dA$ is the electro-magnetic or any $U(1)$ field strength. In reference\cite{Mah09}  we have seen that the current $J^A_{EM}$ is conserved.
Now  we  investigate the  symmetry involved, if any, with this conserved current. 

Here we consider only the globally invariant kinetic part of the spinor $\Psi$ in the Lagrangian $\mbox{$\mathcal{L}$}_{D}$ from reference\cite{Mah07a} in curved space, given by
\begin{eqnarray}
\mbox{$\mathcal{\overline{L}}$}_{D}=\frac{i}{2}\{\overline{\Psi}{}^*\gamma\wedge d\Psi+\overline{d\Psi}\wedge{}^*\gamma\Psi\}\label{eqn:glob}   
\end{eqnarray}It can be verified that under global $SL(2,C)$ transformation on the spinor field and gamma matrices, given by,
\begin{eqnarray}    \Psi\rightarrow\Psi^\prime=S\Psi,\hspace{2mm}\overline{\Psi}\rightarrow\overline{\Psi^\prime}=\overline{\Psi}S^{-1}\hspace{2mm}\mbox{and}\hspace{2mm}\gamma\rightarrow\gamma^\prime=S\gamma S^{-1},
\end{eqnarray}
where $S=\exp(\frac{i}{4}\theta_{ab}\sigma^{ab})$,   $\mathcal{\overline{L}}_{D}$
is an invariant when $\theta_{ab}$ are constant numbers.
Now imposing local invariance in the tangent space where $S=\exp(\frac{i}{4}\theta_{ab}\sigma^{ab})$  $=\exp(\frac{i}{4}\theta_{\mu\nu}\sigma^{\mu\nu})$,  $\theta_{ab}$ are arbitrary infinetesimal tensor-indexed variables, we see that 
\begin{eqnarray}    0&=&\delta\mathcal{\overline{L}}_{D}=\frac{i}{2}\{\overline{\Psi^\prime}{}^*\gamma^\prime\wedge d\Psi^\prime+\overline{d\Psi^\prime}\wedge{}^*\gamma^\prime\Psi^\prime\}\label{eqn:local}\\&{}&-\frac{i}{2}\{\overline{\Psi}{}^*\gamma\wedge d\Psi+\overline{d\Psi}\wedge{}^*\gamma\Psi\},\nonumber\\&=&\frac{i}{2}\overline{\Psi}\{{}^*\gamma\wedge S^{-1}(dS)+(dS^{-1})S\wedge{}^*\gamma\}\Psi,\nonumber\\&=&-\frac{1}{8}\overline{\Psi}\{{}^*\gamma\wedge d(\theta_{\mu\nu} \sigma^{\mu\nu})-d(\theta_{\mu\nu} \sigma^{\mu\nu})\wedge{}^*\gamma\}\Psi,\nonumber\\&{}& (\mbox{neglecting higher order terms of}\hspace{2mm} \theta_{\mu\nu})\nonumber\\&=&-\frac{1}{8}\overline{\Psi}[{}^*\gamma\wedge \{(\bar{\nabla}\theta_{\mu\nu}) \sigma^{\mu\nu}+\theta_{\mu\nu} D(\sigma^{\mu\nu})\}\nonumber\\&{}&-\{(\bar{\nabla}\theta_{\mu\nu}) \sigma^{\mu\nu}+\theta_{\mu\nu} D(\sigma^{\mu\nu})\}\wedge{}^*\gamma]\Psi,\nonumber\\&=&-\frac{1}{8}\overline{\Psi}[{}^*\gamma\wedge (\bar{\nabla}\theta_{\mu\nu}) \sigma^{\mu\nu}-(\bar{\nabla}\theta_{\mu\nu}) \sigma^{\mu\nu}\wedge{}^*\gamma]\Psi,\nonumber\\&=&\frac{1}{8}\bar{\nabla}(\theta_{\mu\nu})\wedge\eta_\alpha\overline{\Psi}(\gamma^\alpha \sigma^{\mu\nu}+ \sigma^{\mu\nu}\gamma^\alpha)\Psi\nonumber\\&=&\frac{1}{8}\bar{\nabla}_\alpha(\theta_{\mu\nu})\overline{\Psi}(\gamma^\alpha \sigma^{\mu\nu}+ \sigma^{\mu\nu}\gamma^\alpha)\Psi\eta,\nonumber\\&=&\frac{i}{4}\bar{\nabla}_\alpha(\theta_{\mu\nu})\overline{\Psi}(\bar{\epsilon}^{\alpha\mu\nu\beta}\gamma_5\gamma_\beta)\Psi\eta=\frac{i}{4}\partial_\alpha(\theta_{\mu\nu})\overline{\Psi}(\bar{\epsilon}^{\alpha\mu\nu\beta}\gamma_5\gamma_\beta)\Psi\eta,\nonumber\\&=&\frac{i}{4}d\theta\wedge(\overline{\Psi}\gamma_5\gamma\Psi)=\frac{i}{4}[d\{\theta\wedge(\overline{\Psi}\gamma_5\gamma\Psi)\}-\theta\wedge d(\overline{\Psi}\gamma_5\gamma\Psi)]\nonumber
\end{eqnarray}
Here the covariant derivative $\bar{\nabla}$ and the matrix valued covariant derivative $D$ have  already been defined to be torsionless. Both act  upon  external and internal indices such that $\bar{\nabla}e^a =0$  and $D\gamma =0$. This gives $D\sigma^{\mu\nu}=0$ and $\bar{\nabla}_\alpha(\theta_{\mu\nu})\bar{\epsilon}^{\alpha\mu\nu\beta}=\partial_\alpha(\theta_{\mu\nu})\bar{\epsilon}^{\alpha\mu\nu\beta}$.  Then arbitrariness of the 2-form $\theta$ implies $\overline{\Psi}\gamma_5\gamma\Psi$ to be a closed 1-form. From reference\cite{Mah09} we know that this 1-form is also exact. Being not a 3-form $\overline{\Psi}\gamma_5\gamma\Psi$ is not  a physical conserved current. But we can introduce an infinitesimal local gauge transformation mediated by the
background $U(1)$ field $F$, given by
\begin{eqnarray} 
   \theta=\kappa\epsilon F
\end{eqnarray}
where $\epsilon$ is an arbitrary infinitesimal scalar field.
With this assumption, equation (\ref{eqn:local}) reduces to
\begin{eqnarray}    0=d\{\kappa\epsilon F \wedge(\overline{\Psi}\gamma_5\gamma\Psi)\}-\kappa\epsilon  d(F\wedge\overline{\Psi}\gamma_5\gamma\Psi)\label{eqn:local1}
\end{eqnarray}and then arbitrariness of $\epsilon$ implies $J^A_{EM}$, a physical conserved
current 3-form, to be  a Noether current\footnote{ We known  that ``The Noether theorem relates symmetries and conserved currents". In the last term of (\ref{eqn:local}), the arbitrariness  of the 2-form $\theta=\frac{1}{2}\theta_{\mu\nu}dx^\mu\wedge dx^\nu=\frac{1}{2}\theta_{ab}e^a\wedge e^b$ gives $d(\overline{\Psi}\gamma_5\gamma\Psi)=0$ i.e. the 1-form  $\overline{\Psi}\gamma_5\gamma\Psi$ is a closed form. In four dimension only closed 3-forms correspond to conserved charges. We know that the axial current $j_5$ is
proportional to the spin current of a fermion\cite{Mie06} and then, considering arbitrariness of the tangent space indexed variables $\theta_{ab}$, we can get from the same term $0=[e^a\wedge e^b\wedge d(\overline{\Psi}\gamma_5\gamma\Psi)]=\bar{\nabla}[e^a\wedge e^b\wedge(\overline{\Psi}\gamma_5\gamma\Psi)]=-\frac{i}{2}\bar{\nabla}[\overline{\Psi}({}^*\gamma\sigma^{ab}+\sigma^{ab}{}^*\gamma)\Psi]$. Hence defining $J^{ab}\equiv \frac{1}{2}e^a\wedge e^b\wedge(\overline{\Psi}\gamma_5\gamma\Psi)$, we may interpret that the 3-form $J^{ab}$ is the matter's angular momentum  current  in the absence of contribution from the orbital part or it is the matter's spin angular momentum current when the corresponding orbital current is separately conserved. This also requires matter's energy momentum stress tensor to be symmetric\cite{Heh97}. It is to be remarked that, by Geroch's theorem\cite{Ger68},  spinors are defined in the local Minkowski space and tetrads are required to define the local tangent  space. So these  two fields are intimately bonded and spinorial matter part of the Lagrangian remains invariant under Lorentz transformation. Here $\theta_{ab}$ are Lorentz transformation parameters in the local tangent space (flat Minkowski space). Since the origin of the inertial frame may be assumed to be attached to the space-time point, we  assume the local inertial frame to be the  rest frame of the matter field attached to this point s.t. the spin angular momentum is the total angular momentum of the spinor field\cite{Kle00}. Our calculation does show the nonappearance of orbital angular momentum in the angular mementum current $J^{ab}$. Here $J^{ab}$ is covariantly conserved and the corresponding conserved charge, bearing local (internal) indices $a,b$, has little physical significances. It is to be noted that,  although there are several promising and useful suggestions, we not only have
no ultimate, generally accepted expression for the energy-momentum and especially for the angular
momentum, but there is not even a consensus in the relativity community on general questions or on the
list of the criteria of reasonableness of such expressions\cite{Sza09}. In view of this it would be better to ponder on some other possible significances connected to the current 3-form $J^{ab}$ or the 1-form $j_5$.
 In this paper we have tried to find a conserved charge, from the axial 1-form $j_5$, bearing no internal (local) index by assuming a background physical electromagnetic field $F=dA$. Since $F$ is a background field it is not arbitrary, its field equations are to be derived by introducing standard terms in the Lagrangian. Arbitrariness of the parameter $\theta$ originates from the scalar parameter $\epsilon$ and then $\theta=\kappa\epsilon F$ is arbitrary with one degree of freedome. Hence the  derivation of the conserved 1-component current $J^A_{EM}$ doesn't contradict Noether's theorem.}.

We know the following theorem, \cite{Rac99,Rac01}:\\
\textit{
Let $(M,g_{\mu\nu})$ be a spacetime associated with a gravity-matter system. Let, furthermore, $\mathfrak{I}$ be an initial hypersurface within an appropriate 
initial value problem. Then there exists a non-trivial Killing vector field $\xi^\mu$
and a gauge potential $A^*_\mu$ so that the Lie derivative $\pounds_\xi A^*_\mu=0$ in a neighbourhood 
$\mathcal{O}$ of $\mathfrak{I}$, if and only if there exists a non-trivial 
initial data set $[\xi^\mu]$, satisfying $\bar{\nabla}_\mu\bar{\nabla}_\nu\xi_\alpha=R_{\nu\alpha\mu\beta}\xi^\beta$,  so that $[\pounds_{\xi}{g}_{\mu\nu}]$ and 
$[\pounds_{\xi}A_\mu]$ vanish identically on $\Sigma$.}\\
Owing to this existence theorem of the Killing vector field $\xi$ there exists, \cite{Wal93,Pad09}, a Noether current 3-form $\mathbf{J}$ and surface density 2-form $\mathbf{Q}$, given by 
\begin{eqnarray} 
   \mathbf{J}&=&d\mathbf{Q}\\
   \mathbf{Q}&=&{}^*d\xi.
\end{eqnarray}Then we can introduce an expression for black hole entropy, \cite{Wal93,Pad09}, given by
\begin{eqnarray}    S_{Noether}\equiv\beta\textit{N}=\beta\int_{\textit{C}}\mathbf{J}=\beta\int_{\partial\textit{C}}\mathbf{Q},\end{eqnarray}here $\beta=\frac{2\pi}{\kappa_0}$, $\kappa_0$ is the surface gravity of the black hole horizion and  $\textit{C}$ is an asymptotically flat hypersurface with ``interior boundary" $\partial\textit{C}$.
 Now we introduce the local $SL(2,C)$ gauge transformation on the spinor field and gamma matrices, mediated by the  surface charge  two form $ d\xi$,  where $\theta$ is
    given by
\begin{eqnarray} 
   \theta=\epsilon d\xi.
\end{eqnarray} Then  equation (\ref{eqn:local}) reduces to 
\begin{eqnarray}    0=d\{\epsilon d\xi \wedge(\overline{\Psi}\gamma_5\gamma\Psi)\}-\epsilon  d(d\xi\wedge\overline{\Psi}\gamma_5\gamma\Psi)\label{eqn:local2}
\end{eqnarray}and  arbitrariness of $\epsilon$ establishes 
\begin{eqnarray}J^A_{KV}\equiv d\xi\wedge\overline{\Psi}\gamma_5\gamma\Psi,\label{eqn:abc247}\end{eqnarray}
 a physical  current 3-form, to be a Noether current.
\section{Discussion} 

By Geroch's theorem\cite{Ger68} we know that - the existence of  the spinor structure is equivalent to the existence of a global field of orthonormal tetrads on the space and time orientable manifold. This requires spinors to be soldered with the tetrads at each space-time point and it justifies the introduction of $\Gamma$ in references\cite{Mah04, Mah05, Mah07a, Mah09} as a frame dependant gauge field. By this way we may assert that ``spinors don't loose their character in the off-shell also". And thus $\Gamma$, being not a frame independant gauge field, corresponds to Noetherian conserved currents  (\ref{eqn:abc147}) or (\ref{eqn:abc247}) which are different from that of standard metric-affine approach\footnote{Here we may compare the currents with equations (5.2.17) and (5.2.19) of \cite{Heh95}. Moreover (\ref{eqn:abc247}) doesn't contradict  Trautman's formula (5.7.10) given in \cite{Heh95}. It can be easily verified that Trautman's formula, in our present approach,  is a sum of two conserved currents. }\cite{Heh95}. Therefore, we see that, in  manifolds having arbitrary background geometry the exterior product of $j_5$ and an $U(1)$ field strength $F$ gives us a gauge invariant conserved current $J^A_{EM}$. Also, following the existance of a Killing vector field $\xi$, there exists another conserved current $J^A_{KV}$. Unlike conserved electric charge in standard model, where the vector current is given by $J\equiv d{}^*F$, conserved charge in the current $J^A_{EM}\equiv\kappa j_5\wedge F$ is associated with the magnetic part of $F$.\footnote{ It may be verified that $J^A_{EM}=\kappa  j_5\wedge F=\kappa F_{ab}J^{ab}$ gives the conserved charge $Q^A_{EM}=\kappa\mathop\int\limits_\Sigma F_{ab}J^{ab}=\kappa\mathop\int\limits_{\partial\Sigma}  A\wedge j_5$ where $\Sigma$ is a spacelike 3-dimensional hypersurface and $\partial\Sigma$ is its 2-dimensional boundary. $Q^A_{EM}$ exists finitely if the dominating terms of the transverse (non-radial) skew-symmetric product of $A_\mu$ and $j_{5\nu}$ behave as $r^{-2}$ at spacial infinity. Considering the anomalous magnatic moment of a  massive spinor\cite{Kak93}, the conserved charge $Q^A_{EM}$ may be associated with the energy current of the spin-magnet in the magnetic field. } In a similar manner the current $J^A_{KV}$ is associated with the magnetic part of $d\xi$ where as Noetherian entropy is connected to its electric part\cite{Wal93,Pad09,Wal94}. In this sense, the conserved charges of $J^A_{EM}$ and $J^A_{KV}$ are connected to monopole charges. Further study is in progress to explore other significances, more specific in nature, which are to be borne by these conserved currents.
\bibliographystyle{my-h-elsevier}

\end{document}